\documentclass[10pt,a4paper,twocolumn]{article}
\usepackage{f1000_styles}

\usepackage[numbers]{natbib}
\usepackage{aas_macros}

\begin{document}
\pagestyle{fancy}

\title{Atacama Large Aperture Submillimeter Telescope (AtLAST) Science: Planetary and Cometary Atmospheres}

\author[1,2]{Martin A. Cordiner}
\author[3]{Alexander E. Thelen}

\author[4,5]{Thibault Cavali{\'e}}
\author[6]{Richard Cosentino}
\author[7]{Leigh N. Fletcher}
\author[8]{Mark Gurwell}
\author[3]{Katherine de Kleer}
\author[9]{Yi-Jehng Kuan}
\author[5]{Emmanuel Lellouch}
\author[10]{Arielle Moullet}
\author[11]{Conor Nixon}
\author[12]{Imke de Pater}
\author[13]{Nicholas A. Teanby}

\author[14]{Bryan Butler}
\author[1]{Steven Charnley}
\author[1]{Stefanie Milam}
\author[5]{Raphael Moreno}
\author[15]{Mark Booth}
\author[15]{Pamela Klaassen}
\author[16]{Claudia Cicone}
\author[17]{Tony Mroczkowski}
\author[18,19,20,21]{Luca Di Mascolo}
\author[22,23]{Doug Johnstone}
\author[17]{Eelco van Kampen}
\author[24,25]{Minju M. Lee}
\author[26,27]{Daizhong Liu}
\author[28]{Thomas J. Maccarone}
\author[26,29]{Am\'{e}lie Saintonge}
\author[30]{Matthew Smith}
\author[16,31]{Sven Wedemeyer}

\affil[1]{Astrochemistry Laboratory, Code 691, NASA Goddard Space Flight Center, Greenbelt, MD 20771, USA}
\affil[2]{Department of Physics, Catholic University of America, Washington, DC 20064, USA}
\affil[3]{Division of Geological and Planetary Sciences, California Institute of Technology, Pasadena, CA 91125, USA}
\affil[4]{Laboratoire d’Astrophysique de Bordeaux, Univ. Bordeaux, CNRS, B18N, all{\'e}e Geoffroy Saint-Hilaire, 33615, Pessac, France}
\affil[5]{LESIA, Observatoire de Paris, PSL Research University, CNRS, Sorbonne Universit{\'e}, Universit{\'e} de Paris, 5 place Jules Janssen, 92195 Meudon, France}
\affil[6]{Space Telescope Science Institute, Baltimore, MD 21218, USA}
\affil[7]{School of Physics and Astronomy, University of Leicester, University Road, Leicester LE1 7RH, UK}
\affil[8]{Center for Astrophysics, Harvard \& Smithsonian, Cambridge, MA 02138, USA}
\affil[9]{National Taiwan Normal University, Taipei 116, Taiwan, ROC}
\affil[10]{National Radio Astronomy Observatory, 520 Edgemont Road, Charlottesville, VA 22903, USA}
\affil[11]{Planetary Systems Laboratory, Code 693, NASA Goddard Space Flight Center, Greenbelt, MD 20771, USA}
\affil[12]{Departments of Astronomy and Earth and Planetary Science, University of California Berkeley, Berkeley, CA 94720, USA}
\affil[13]{School of Earth Sciences, University of Bristol, Wills Memorial Building, Queens Road, Bristol BS8 1RJ, UK}
\affil[14]{National Radio Astronomy Observatory, Socorro, NM 87801, USA}

\affil[15]{UK Astronomy Technology Centre, Royal Observatory Edinburgh, Blackford Hill, Edinburgh EH9 3HJ, UK}
\affil[16]{Institute of Theoretical Astrophysics, University of Oslo, PO Box 1029, Blindern 0315, Oslo, Norway}
\affil[17]{European Southern Observatory (ESO), Karl-Schwarzschild-Strasse 2, Garching 85748, Germany}
\affil[18]{Laboratoire Lagrange, Université Côte d'Azur, Observatoire de la Côte d'Azur, CNRS, Blvd de l'Observatoire, CS 34229, 06304 Nice cedex 4, France}
\affil[19]{Astronomy Unit, Department of Physics, University of Trieste, via Tiepolo 11, Trieste 34131, Italy}
\affil[20]{INAF -- Osservatorio Astronomico di Trieste, via Tiepolo 11, Trieste 34131, Italy}
\affil[21]{IFPU -- Institute for Fundamental Physics of the Universe, Via Beirut 2, 34014 Trieste, Italy}
\affil[22]{NRC Herzberg Astronomy and Astrophysics Research Centre, 5071 West Saanich Rd, Victoria, BC, V9E 2E7, Canada}
\affil[23]{Department of Physics and Astronomy, University of Victoria, Victoria, BC, V8P 5C2, Canada}

\affil[24]{Cosmic Dawn Center (DAWN), Denmark}
\affil[25]{DTU-Space, Technical University of Denmark, Elektrovej 327, DK2800 Kgs. Lyngby, Denmark}
\affil[26]{Max-Planck-Institut f\"{u}r extraterrestrische Physik, Giessenbachstrasse 1 Garching, Bayern, D-85748, Germany}
\affil[27]{Purple Mountain Observatory, Chinese Academy of Sciences, 10 Yuanhua Road, Nanjing 210008, China}
\affil[28]{Department of Physics \& Astronomy, Texas Tech University, Box 41051, Lubbock, TX 79409, USA }
\affil[29]{Department of Physics and Astronomy, University College London, Gower Street, London WC1E 6BT, UK}

\affil[30]{School of Physics \& Astronomy, Cardiff University, The Parade, Cardiff, CF24 3AA, UK}
\affil[31]{Rosseland Centre for Solar Physics, University of Oslo, Postboks 1029 Blindern, N-0315 Oslo, Norway}

\maketitle
\thispagestyle{fancy}

\clearpage
\begin{abstract}

The study of planets and small bodies within our Solar System is fundamental for understanding the formation and evolution the Earth and other planets. Compositional and meteorological studies of the giant planets provide a foundation for understanding the nature of the most commonly observed exoplanets, while spectroscopic observations of the atmospheres of terrestrial planets, moons, and comets provide insights into the past and present-day habitability of planetary environments, and the availability of the chemical ingredients for life. While prior and existing (sub)millimeter observations have led to major advances in these areas, progress is hindered by limitations in the dynamic range, spatial and temporal coverage, as well as sensitivity of existing telescopes and interferometers. Here, we summarize some of the key planetary science use cases that factor into the design of the Atacama Large Aperture Submillimeter Telescope (AtLAST), a proposed 50-m class single dish facility: (1) to more fully characterize planetary wind fields and atmospheric thermal structures, (2) to measure the compositions of icy moon atmospheres and plumes, (3) to obtain detections of new, astrobiologically relevant gases and perform isotopic surveys of comets, and (4) to perform synergistic, temporally-resolved measurements in support of dedicated interplanetary space missions. The improved spatial coverage (several arcminutes), resolution ($\sim1.2''-12''$), bandwidth (several tens of GHz), dynamic range ($\sim10^5$) and sensitivity ($\sim1$~mK\,km\,s$^{-1}$) required by these science cases would enable new insights into the chemistry and physics of planetary environments, the origins of prebiotic molecules and the habitability of planetary systems in general.

\end{abstract}

\section*{\color{OREblue}Keywords}

Planets; Comets; Planetary atmospheres; Spectral lines; Spectral imaging; Submillimeter; Instrumentation

\clearpage
\pagestyle{fancy}

\section*{Plain Language Summary}
Our present understanding of what planets and comets are made of, and how their atmospheres move and change, has been greatly influenced by observations using existing and prior telescopes operating at wavelengths in the millimeter/submillimeter range (between the radio and infrared parts of the electromagnetic spectrum), yet major gaps exist in our knowledge of these diverse phenomena. Here, we describe the need for a new telescope capable of simultaneously observing features on very large and very small scales, and covering a very large spread of intrinsic brightness, in planets and comets. Such a telescope is required for mapping storms on giant planets, measuring the compositions of the atmospheres and plumes of icy moons, detecting new molecules in comets and planetary atmospheres, and to act as a complement for measurements by current and future interplanetary spacecraft missions. We discuss the limitations of currently-available millimeter/submillimeter telescopes, and summarize the requirements and applications of a new and larger, more sensitive facility operating at these wavelengths: the Atacama Large Aperture Submillimeter Telescope (AtLAST).

\section*{Introduction}

Since the latter half of the 20th century, (sub)millimeter ($\sim0.3-3$ mm; 100--950 GHz) telescope facilities have been increasingly employed for the study of planetary surfaces, atmospheres, and ring/moon systems. Brightness temperature measurements can be readily derived for large solar system objects using traditional single-dish radio facilities, while the high spectral resolution combined with increasing bandwidth and sensitivity of modern (sub)millimeter wave telescopes and interferometers has enabled pioneering spectroscopic surveys and spatial-spectral mapping of planetary bodies and comets. The measurement of thermal emission in the shallow subsurfaces of natural satellites, small bodies, and terrestrial planets fills the gap between radar sounding of the deep (\textgreater1 km) surface and the near surface temperatures derived from infrared observations. 

The (sub)millimeter wavelength regime also provides access to rotational transitions from gas-phase molecules in planetary atmospheres, the observation of which leads to robust, spectroscopic molecular detections as well as abundance, temperature, and wind measurements with well-defined, and often high, degrees of quantitative accuracy. The available molecular species tend to be dominated by the more abundant reactive elements in our solar system (C, H, N, O, P and S), and therefore include hydrocarbons, oxides, hydrides, nitriles and other members of the class of organic molecules associated with terrestrial biology. 

The Atacama Large Aperture Submillimeter Telescope (AtLAST) is proposed as a more powerful, ground-based, single-dish submillimeter facility to begin operating in the 2030s \cite{kla20,ram22,mro23,mro24}. Combining improved sensitivity, angular resolution and field-of-view, AtLAST would open new frontiers in our understanding of planetary atmospheres and surfaces, as well as providing more detailed characterization of the gases produced by comets.  More sensitive access to the long-wavelength region of the spectrum is also crucial for providing supporting observations and large-scale context for the more focused measurements performed by interplanetary spacecraft missions. In particular, the availability of improved mapping capabilities, wider instantaneous bandwidth, and higher spatial resolution made possible by AtLAST would result in improvements in our ability to characterize and monitor dynamic and complex physical phenomena, such as winds and storms, as well as chemical processes (revealed by rotational spectroscopy), occurring in the atmospheres and on the surfaces of planetary bodies. 

The study of `local' planets in our Solar System also provides a fundamental baseline for interpreting the properties of the vast number of exoplanets that are now known to exist around other stars throughout the galaxy \cite{zhu21}. The paradigm within which we understand the environmental conditions, climate and possible habitability of exoplanets exists by virtue of the active development of an extensive knowledge-base on atmospheric (and bulk) chemical inventories and physical processes occurring in the planets, moons, and other minor bodies of our Solar System. In this era of rapid discovery, it is thus more important than ever to develop and build new telescope facilities such as AtLAST, to continue to advance our understanding of the compositions, spatial and temporal features, and evolutionary processes occurring on the bodies within our Solar System.

\subsection*{Prior/existing instruments and results}

Single-dish radio/submillimeter facilities provide a wealth of unique and complementary observations to spacecraft and facilities at other wavelengths. The Owens Valley Radio Observatory (OVRO), Institut de Radioastronomie Millimétrique (IRAM) 30-m telescope, Green Bank Observatory (GBO), Arecibo radar facility, among others, were previously used to study the brightness temperatures of the Giant planets, Mars, Venus, and large satellites. They enabled study of the deep atmospheric composition of Gas and Ice Giants, characterization of the near subsurface properties of the Galilean Satellites, as well as compositional and dynamical studies of terrestrial planet and satellite atmospheres. These were complemented by early interferometric observations by the Very Large Array (VLA), Berkeley-Illinois-Maryland Association (BIMA) array, Submillimeter Array (SMA), and Atacama Large Millimeter/submillimeter Array (ALMA), enabling higher sensitivity and angular resolution studies \cite{dep82,dep84,muh86,alt88,lel94,won96,mar02,gur04,mor05,dep19a,dep21a,dek21a, dek21b}. 

Cometary science --- the study of our Solar System’s oldest, yet most pristine materials --- has benefited hugely from advances in single-dish millimeter-wave instrumentation. The first detections of key organics HCN and CH$_3$OH in cometary comae were obtained through observations using the National Radio Astronomy Observatory (NRAO) 11-m and IRAM 30-m telescopes in the 3~mm and 2~mm bands, respectively \cite{hue74,boc94}. Millimeter-wave spectroscopy continues to provide the primary method for remotely detecting and characterizing new molecules (including complex organic molecules) in cometary comae \cite{biv15}, while also providing crucial insights into the origins of our solar system’s primitive materials via detailed studies of molecular isotopic ratios \cite{har11,biv16}. ALMA is currently revolutionizing the study of comets at (sub)millimeter wavelengths, revealing the coma and nucleus outgassing sources in unprecedented detail from the ground \cite{cor14,cor17,cor23}, yet snapshot maps of the largest coma scales remain out of reach with present facilities. A new, highly-sensitive single-dish (sub)millimeter telescope with wide-field imaging capabilities will help advance coma mapping studies and help break new ground in our understanding of the chemical composition of comets. 

Observations at (sub)millimeter wavelengths have enabled the detection and mapping of new molecular species in the atmospheres of various planets in our solar system. CO and HCN were first detected on Neptune in the early 1990s using the James Clerk Maxwell Telescope (JCMT) and CalTech Submillimeter Observatory (CSO) telescopes \cite{mar93}, with abundances  $\sim1000$ times higher than predicted from thermochemical models. Similarly, CO, HCN, and HNC were detected on Pluto with recent ALMA observations \cite{lel17}, and a variety of trace gases have been discovered on Titan using millimeter-wave spectroscopy \cite{bez93,cor15,cor19,pal17,nix20,the20}.

Sub-millimeter observations can also be used to probe gas and ice giant interiors. For example, Neptune's very high observed CO abundance in the stratosphere and troposphere, combined with thermochemical models, has been used to argue for an ice-dominated interior \cite{lod94,lel05,cav17}. However, a comet impact in the last few hundred years, as suggested by (sub)millimeter observations of stratospheric CO \cite{lel05} and CS \cite{mor17}, would imply significant external flux into Neptune's atmosphere. Therefore, a rock dominated interior is also plausible \cite{tea20} and more consistent with D/H ratio measurements if the interior is well mixed \cite{feu13}.

The Herschel Space Observatory provided numerous demonstrations of the power of (sub)millimeter single-dish heterodyne spectroscopy for characterizing planetary atmospheres, producing the first detection of water in the extended gaseous torus around Enceladus \cite{har11b}. The Herschel Heterodyne Instrument for the Far-Infrared (HIFI) instrument also obtained the first detection of hydrogen isocyanide (HNC) in Titan’s atmosphere \cite{mor11}. Stratospheric H$_2$O measurements with HIFI have been used to demonstrate the cometary origin of water in Jupiter's stratosphere \cite{cav13} and to constrain the background flux of interplanetary dust particles into Uranus' and Neptune's atmospheres \cite{tea22}. \citet{fletch12} used the Spectral and Photometric Imaging Receiver (SPIRE) instrument on Herschel to constrain the vertical distributions of phosphine, ammonia and methane on Saturn, as well as providing measurements of stratospheric water and precise upper limits for a range of exotic compounds including halides.

In general, the improvements in (sub)millimeter wave sensitivity and mapping capabilities ushered in by ALMA have led to major strides in our understanding of the compositions and dynamical states of planetary atmospheres, as well as satellite and small body surfaces \citep{cord14, cord15, cord18, cord19, cord20, the18, the19a, the19b, the20, the22, the24a, mor17, iino18, dep19a, dek21a, dek21b, toll19, toll21, cav21, cav23, ben21, ben22, cam23, carrg23}. However, the lack of simultaneous, high-sensitivity total power observations hinders ALMA's ability to characterize localised/dynamical phenomena such as storms and plumes on rotating planetary surfaces, due to temporal smearing.  A new, single-dish telescope with improved total power sensitivity and wide-field mapping capabilities is therefore required.

\begin{figure*}
\centering
\includegraphics[width=0.7\textwidth]{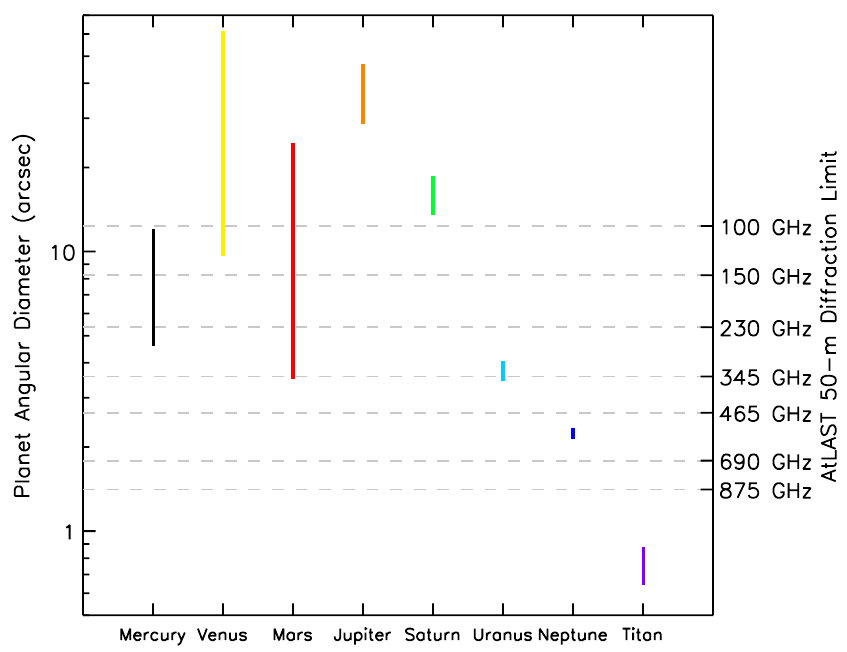}
\caption{\label{fig:res} Angular diameters of major Solar System bodies and Saturn's largest moon, Titan. The vertical extents of the coloured bars for each body represent their range of angular sizes due to differing geocentric distances throughout the year. The second y-axis shows the angular resolution a 50 m single dish facility would achieve for a range of representative frequencies.}
\end{figure*}

Instantaneous (snapshot) mapping over the entire area of a body is critical for detailed studies of the rapidly rotating and evolving atmospheres of solar system objects. Such studies have historically been the realm of interferometry, though the lack of short baselines in a typical interferometric array leads to spatial filtering of the resulting images, which can preclude reliable measurements of extended emission from larger bodies such as the giant planets and extended cometary comae. Further, the re-positioning of antennas to larger configurations (such as the extended baselines of ALMA) may impede mapping of the giant planets during transient events, such as impacts and storms: impact events tell us both about the population, bulk properties, and volatile composition of potential impactors in the outer Solar System, while storms provide us access to the chemical composition deep below the topmost clouds, and reveal how dramatic meteorology evolves in non-terrestrial and cold environments. Combined with their large angular scales (Figure \ref{fig:res}) and total flux, the planets present challenging targets for observation of small brightness temperature variations or weak spectral lines against a strong background continuum at (sub)millimeter wavelengths, and therefore place stringent demands on the stability, calibration, and dynamic range of heterodyne telescope instruments. The availability of a new, cutting edge single-dish (sub)millimeter facility at a dry, high-altitude site will help to fill this gap, allowing us to address important outstanding questions in planetary and cometary science as outlined in this article.

\subsection*{The Need for a Larger Aperture, Multi-Beam, Single Dish (Sub)Millimeter Facility}

\textbf{1. Improved Sensitivity and Dynamic Range for Molecular Detections}

The spectral and imaging dynamic range limitations of currently-leading facilities such as ALMA hinder our ability to detect and map new chemical species in planetary atmospheres. ALMA's nominal spectroscopic dynamic range (the ratio of strongest to weakest signals in the spectrum --- typically expressed as the line-to-continuum intensity ratio), is $\sim1000$ at 1.3~mm, but falls to 150--400 at shorter wavelengths ($\sim0.7$ mm) using standard calibration procedures. Such limitations make it difficult to reliably measure weak spectral lines and detect new molecules in the (sub)millimeter range. This is a particular problem for planetary targets that have a relatively bright (sub-)mm continuum.  The maximum achievable dynamic range is dictated by the ability to accurately calibrate (or flatten) the spectroscopic baseline, which, for interferometers like ALMA, requires the observation of a point-like calibration source with a strong, line-free continuum. Improved calibration accuracy of a single-dish (`total power') facility such as AtLAST could be achieved through observations of brighter, more extended, spectrally featureless solar system objects (such as the Moon or Mercury). This would enable more precise calibration of the intrinsic receiver bandpass.

Advances in instrument design and optics since the construction of present-generation (sub)millimeter facilities such as the IRAM 30-m and JCMT will further facilitate the acquisition of spectra at high dynamic range, allowing the detection of new, spatially distributed trace gases in planetary atmospheres, such as complex organic molecules and other (perhaps biologically relevant) species. For optimal science return on AtLAST, a spectroscopic dynamic range $\sim10^5$ across large bandwidths should be the goal (see Science Case ii). To realize this objective, close attention should be paid to achieving the flattest possible spectral bandpass, through careful choice of optical design, receiver components, and optimal calibration strategies. To this end, it will also be important to mitigate internal reflections and standing waves between the telescope components as much as possible.

Cometary comae are highly diffuse and extended objects. Their molecular excitation is mostly governed by a combination of thermal and fluorescent processes, leading to rotational emission lines with peak intensities in the (sub)millimeter band. Nevertheless, molecular column densities remain low, so that detection of new molecules and mapping of known species can be challenging in all but the brightest comets. Studies of cometary chemical compositions therefore benefit in proportion to the available spectral line sensitivity; ideally, AtLAST would provide a factor of at least a few improvement compared with existing facilities. This can be achieved through a combination of $\sim2\times$ smaller beam size for a given frequency (focusing in on the denser, inner-coma), combined with lower receiver temperatures, and reduced sky opacities compared with the IRAM 30-m, which is the current state-of-the art instrument for total power mm-wave  spectroscopy of comets. Our goal sensitivity for new spectroscopic detections of cometary molecules with AtLAST should be $\sim1$ mK (per km\,s$^{-1}$ of bandwidth) in $\lesssim8$~h observing time (see Science Case iv).\\

\textbf{2. High Spatial Resolution}

The relatively low spatial resolution of current single-dish (sub)millimeter facilities has limited the potential for mapping of planetary atmospheres and surfaces. The angular diameters of the major planets (and Titan) are shown in Figure \ref{fig:res}, demonstrating the spread in their apparent sizes over time due to orbital motions. For comparison, the diffraction-limited angular resolution (primary beam FWHM) for a 50~m diameter antenna at a variety of frequencies is also shown. The maximum number of resolution elements within the area of the major Solar System bodies as a function of these frequencies (and wavelengths) are given in Table \ref{tab:beams}. These data demonstrate that a 50-m class single-dish (sub)millimeter telescope can readily resolve Venus, Mars, Jupiter and Saturn ---- often, throughout their orbits, in a number of frequency ranges --- while the Ice Giants (Uranus and Neptune) can be marginally resolved at terahertz frequencies. Titan remains smaller than $1''$ throughout its year, but is still an important target for total power spectroscopy with AtLAST, since a 50~m single dish would allow high-frequency observations without severe beam dilution (particularly above $\sim800$ GHz).

Large planets such as Jupiter and Saturn are a particular challenge for interferometers like ALMA, NOEMA and SMA as they possess large angular structures that are resolved out at long baselines. As shown in Figure \ref{fig:res} and Table \ref{tab:beams}, a large-aperture, single-dish telescope would enable the observation of planetary objects and surface/atmospheric features that span a wider range of spatial scales. Providing planetary observers with a consistent range of angular resolutions \textit{at all times}, would facilitate studies of temporal phenomena, contemporaneous observations with other facilities, and support for spacecraft measurements. This latter task can prove difficult for interferometers, which tend to have variable configurations (and thus, variable angular sensitivities) throughout the observing schedule. Further, a single-dish instrument with focal-plane receiver array at least an arcminute in diameter would permit mapping of both large and small-scale planetary features. This is particularly valuable for moderately well mixed gases that may only vary on global scales. 

A small beam size of $\lesssim5''$ is required to resolve the majority of planets in our Solar System, which is achievable at frequencies \textgreater250 GHz for a 50~m facility. Additionally, a small beam would help minimize beam dilution for the small ($\sim0.1''$--$2''$ diameter) bodies such as the icy moons of Jupiter, and will be required to ensure Saturn's moon Enceladus can be reliably separated from Saturn's bright continuum emission (see Science Case iii).\\


\begin{table*}
   \caption{\label{tab:beams} Maximum Number of Beams Per Planet Area for 50 m Diffraction-Limited Aperture}
   \centering
   \begin{tabledata}{cccccccccc}
    \header Wave. & Freq. & Ang. Res. & Mercury & Venus & Mars & Jupiter & Saturn & Uranus & Neptune \\
   \header (mm) & (GHz) & ($"$) & & & & & & & \\
3.00 & 100 & 12.4 & 9 & 239 & 38 & 139 & 22 & 1 & \\
2.00 & 150 & 8.2 & 14 & 359 & 57 & 208 & 33 & 2 & \\
1.30 & 230 & 5.4 & 21 & 551 & 88 & 319 & 51 & 2 & \\
0.87 & 345 & 3.6 & 32 & 826 & 131 & 478 & 76 & 4 & 1 \\
0.64 & 465 & 2.7 & 43 & 1113 & 177 & 644 & 103 & 5 & 2 \\
0.43 & 690 & 1.8 & 63 & 1652 & 263 & 956 & 153 & 7 & 2 \\
0.34 & 875 & 1.4 & 80 & 2095 & 333 & 1212 & 193 & 9 & 3 \\
0.30 & 1000 & 1.2 & 91 & 2394 & 381 & 1385 & 221 & 11 & 3 \hspace*{-2mm}
   \end{tabledata} 
 \end{table*}

\textbf{3. Instantaneous Multi-beam Spectral Mapping}

AtLAST has the potential to capitalize on recent advances in heterodyne multiplexing and multi-beam receiver technology to produce unprecedented spectro-spatial sub-mm maps of the larger planets, as well as cometary comae, the latter of which often extend over hundreds of arcseconds on the sky. The ability to map in two spatial dimensions the chemical distributions and dynamical motions of Solar System objects at a given instant in time is crucial for properly characterizing short-term phenomena such as planetary impacts and storms, and cometary jets and outbursts. Spatial resolution of a few arcseconds over a field of view up to a few arcminutes, with $\sim10$--30 beams across a single axis (for example, in a square or hexagonal, $\sim25\times25$ pixel array), at a spectral resolution $\sim0.1$~km\,s$^{-1}$, would allow detailed imaging and Doppler studies of molecular species in planetary and cometary atmospheres to study the combined effects of chemical and dynamical processes in these bodies.

\section*{Science Case i: Giant Planet Atmospheres}

The complex, extended atmospheres of the Gas and Ice giant planets provide interesting, yet challenging targets for remote observations due to (1) small-scale vertical (radial) variations in temperature and composition, (2) the brightness of their continuum emission (particularly in contrast to trace atmospheric species), and (3) their high atmospheric pressures and fast rotation speeds, which can significantly broaden the spectral lines. Millimeter/submillimeter facilities provide powerful probes of the composition, structure, and dynamics of the tropospheres and stratospheres of giant planets through the combination of rotational emission/absorption line measurements, studies of highly pressure-broadened (i.e. \textgreater1 GHz wide) pseudo-continuum absorption features, and spatial mapping of brightness temperature variations. 

The continuum emission from giant planets is dominated by the temperature structure and deep abundances of H$_2$, He, NH$_3$, H$_2$S, PH$_3$, CH$_4$, and other gases, which are difficult properties to observe directly through remote sensing and often require \textit{in situ} measurements through interplanetary probes \citep{won04, mou14, mou18}. With the advancement of larger aperture and interferometric facilities, the spatial variability of the giant planet continuum emission can be mapped to better understand the connection between deeper atmospheric composition and dynamical activity, and will help to contextualize spacecraft observations including those made by \textit{Juno} and the recently launched \textit{JUpiter Icy Moons Explorer (JUICE)} mission \citep{fletch23}. The development of a larger-aperture, single-dish telescope (AtLAST) will further help to improve our understanding of gas and ice giant planet atmospheres -- and by extension, exoplanets with similar sizes and compositions -- in several key areas:

\begin{itemize}
   \item The spatial and temporal variation of temperature structure throughout the troposphere and stratosphere, which relates to the global and local circulation, dynamics and composition of the atmosphere. \vspace{-3mm}
    \item Variability of atmospheric winds and vertical wind shear on short-term and seasonal timescales, and how that relates to climatological and other forcing mechanisms. \vspace{-3mm}
   \item The composition and distribution of trace species throughout the stratosphere and troposphere. \vspace{-3mm}
   \item The changes in chemical abundances over short and long timescales, and their connection to seasonal and transient events such as storms, aurorae and infall from space (cometary impacts, micrometeorites and dust).
\end{itemize}

\subsection*{Dynamics}

\begin{figure*}
\centering
\includegraphics[width=\textwidth]{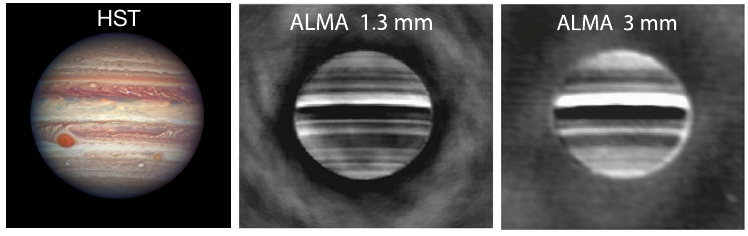}
\caption{\label{fig:jup} HST (left) and longitude-smeared ALMA (center, right) observations of Jupiter from \citet{dep19a} (used with permission). The ALMA observations show data after the subtraction of a limb-darkened disk model, enabling the high contrast ($\lesssim10$ K) differences in brightness temperature between Jupiter's zonal structure to be easily observed.}
\end{figure*}

The dynamical state of planetary atmospheres -- often described by their temperature structure and wind fields -- allows for the characterization of their atmospheric circulation, energy budget, and seasonal variability. Previous studies at long wavelengths have measured brightness temperature and compositional variations in the giant planets to assess the influence of seasonal changes in insolation, meteorological activity, and transient events (such as impacts) throughout their tropospheres and stratospheres, which complement studies in the optical and infrared. The temperature and circulation of Neptune's troposphere and stratosphere have been studied using the the Combined Array for Research in Millimeter-wave Astronomy (CARMA), VLA, and ALMA \citep{lus13a, lus13b, iino18, toll19, toll21} through observations of CO and continuum features. Similarly, the NH$_3$ and brightness temperature distribution on Jupiter from VLA and ALMA observations has been used to contextualize storm outbreaks and observations with the \textit{Juno} spacecraft \citep{dep19a, dep19b, moe23}. Figure \ref{fig:jup} shows the comparison of ALMA observations of Jupiter at 1.3 and 3.0 mm to a visible wavelength composite from the Hubble Space Telescope (HST), allowing for the comparison of brightness temperatures at multiple pressure levels from various zonal features. At frequencies \textgreater700 GHz, a 50 m single-dish facility with large focal plane receiver array will be able to provide comparable latitudinal coverage to ALMA Band 3 observations (Figure \ref{fig:jup}, right panel) in a single integration. Sensitivity $\lesssim1$ K over short integration times (minutes) would allow for the precise determination of brightness temperature variations compared to the bright continuum emission while avoiding longitudinal smearing of localized features due to planetary rotation and winds; higher sensitivity and dynamic range considerations pertain mostly to spectral line surveys, as discussed below. Wide bandwidth (\textgreater a few GHz) settings enable the pressure-broadened rotational transitions (e.g., CO at $\sim115$, 231, 346, 461, 576, 691, 807, and 922 GHz) and pseudo-continuum features (e.g., H$_2$S at $\sim169$, 408, and 736 GHz; PH$_3$ at $\sim267$ GHz; and NH$_3$ at $\sim572$ GHz) to be modeled so the deep temperature and composition can be characterized, which complements observations in the IR down to $\sim1$ bar (e.g., \citealp{fletch20}). Observing the giant planets multiple times throughout their extensive orbital periods will improve our understanding of the influence of seasonal variability on the circulation and dynamical state of the atmosphere and allow for comparison to seasonal photochemical models \citep{mos05, mos20, hue16, hue18}.

In contrast to the meridional circulation of the atmosphere, which is too slow to readily detect using spectral line shifts, the direct measurement of horizontal zonal atmospheric winds is possible from the Doppler shifts of molecular rotational transitions at high ($\sim0.1$ km s$^{-1}$) spectral resolution. Wind speeds of the giant planets have typically been inferred directly from cloud tracking techniques (e.g. through Voyager imaging; \citealp{lim91}), which rely on correlation analyses of small-scale features observed over multiple rotations. Zonal wind speeds can also be inferred  through infrared measurements of the thermal wind shear \citep{fletch20b}. Direct observation of spectral line (e.g., HCN at $\sim89$, 265, 355, and 709 GHz) Doppler shifts are required in order to obtain precise wind speed measurements at altitudes \textit{above} the clouds, avoiding the problems associated with changes in the shapes of resolved atmospheric features over multiple rotations. 

\begin{figure*}
\centering
\raisebox{1mm}{\includegraphics[width=0.52\textwidth]{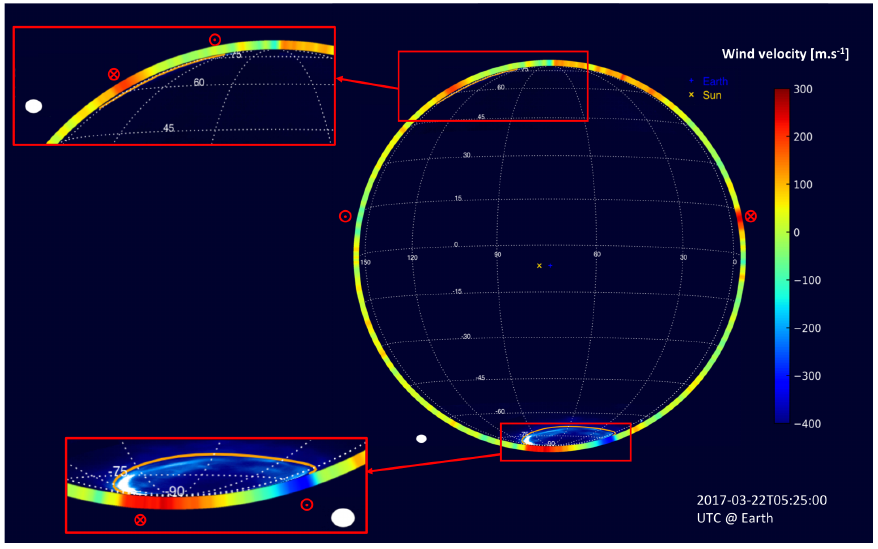}}
\includegraphics[width=0.29\textwidth]{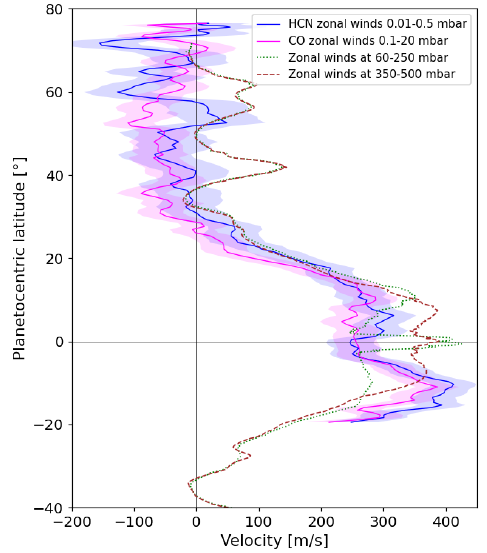}
\includegraphics[width=0.17\textwidth]{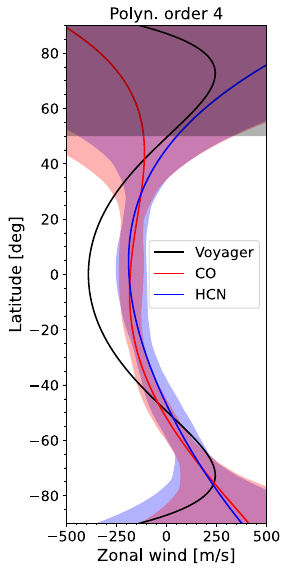}
\caption{\label{fig:gp_winds}(Left) Jupiter zonal wind velocities derived from ALMA observations of HCN at $\sim1''$ resolution (color map); ultraviolet auroral emission at the south pole is also shown \citep{cla09, cav21}. (Center) Zonal wind speeds as a function of latitude in Saturn's stratosphere from \citet{ben22}, compared to winds from Cassini imaging. (Right) Comparison of ALMA wind speed measurements of Neptune from CO and HCN emission lines compared to Voyager cloud tracking measurements \citep{carrg23}. Images used with permission from their respective copyright holders.}
\end{figure*}

Observations of strong stratospheric rotational transitions of CO and HCN using ALMA has resulted in wind measurements for all giant planets apart from Uranus \citep{cav21, ben22, carrg23}. Figure \ref{fig:gp_winds} shows examples of winds derived from ALMA observations of Jupiter, Saturn, and Neptune compared to cloud tracking and auroral measurements. Measuring planetary wind speeds as a function of latitude and at multiple altitudes requires high spectral and spatial resolution combined with a large field of view, which is enabled by heterodyne and (sub)millimeter instrumentation. While the variability in spatial resolution of the giant planets (Figure \ref{fig:res}, Table \ref{tab:beams}) precludes complete latitudinal coverage for all planets at all frequencies, even hemispheric comparisons of Doppler wind measurements over time provide interesting results --- see, for example, the abrupt change in Titan's stratospheric jet \citep{lel19, cord20}. As a single-dish facility would not be confined to time-variable configuration (resolution) constraints, episodic changes in wind speeds and wave propagation can be investigated frequently, and over irregular timescales as needed.

Jupiter and Saturn exhibit large-scale weather events where dramatic changes in the planet's visible appearance are accompanied by energetic changes in cloud color and morphology. On Jupiter, such events periodically take place at low latitudes \citep{fletch17a}, in the typically white band know as the Equatorial Zone  ($\pm7^\circ$ latitude), and in the adjacent, darker North and South Equatorial Belts between $\pm7-18^\circ$ latitude. Each region undergoes changes in its normal cloud color and morphology, exhibiting waves, eddies, and wakes, initially beginning over a small local region and expanding to encompass the whole region for some time before returning back to its nominal state. These events on Jupiter happen on a timescale of between 2--18 years, with little correlation to adjacent regional activity, season, or orbital position \citep{antu18, antu23}. On Saturn, however, the northern hemisphere experiences a large convective storm event approximately every 30 years. The previous `Great Saturn Storm' in 1990 occurred prior to the advent of the ALMA telescope, spanning $\sim15^\circ$ latitude and eventually encircling the whole planet. This event led to significant stratospheric effects that could have been readily characterized in the (sub)millimeter \citep{fletch12}. Additionally, `Great White Spots' have previously been observed with Herschel \citep{cav12}. The sizes of these types of spectacular atmospheric events ($\sim2-5"$) would make them observable by AtLAST. As extensive meteorological events offer a window into the planets' deeper atmospheres, contemporaneous (sub)millimeter observations (which can probe different altitudes by highly resolving line profiles and measuring multiple rotational transitions), provide complementary information to other facilities (e.g., HST, JWST, the \textit{Juno} Microwave Radiometer instrument). Since a large, single-dish facility would provide adequate spatial resolution to discern between the $\sim1-5"$ zonal regions on the giant planets (see Table \ref{tab:beams}, Figure \ref{fig:jup}) over their orbital period, strategic and regular observations by AtLAST would serve as an excellent resource to investigate these spectacular episodic events and provide global context for other ground-based and spacecraft measurements.

\subsection*{Chemistry}

\begin{figure*}
\centering
\includegraphics[width=0.95\textwidth]{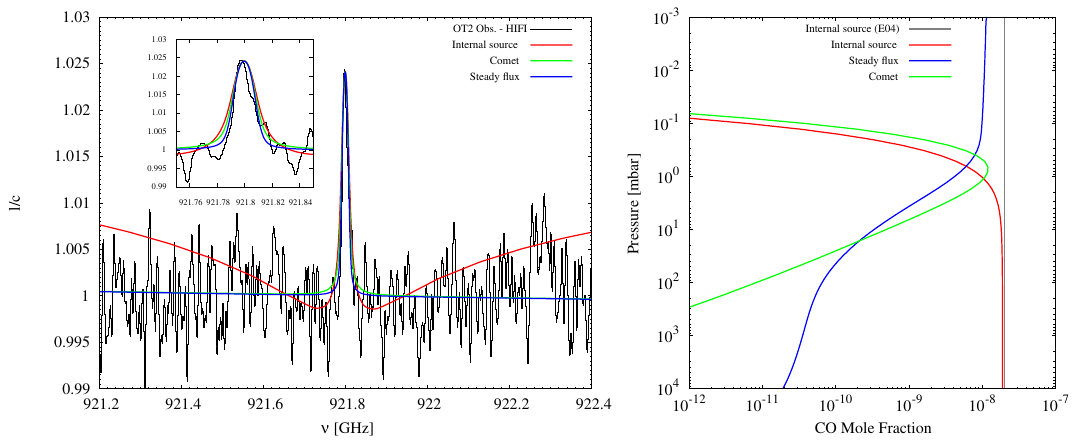}
\caption{\label{fig:ur_co} Figure from \citet{cav14} (used with permission) showing Herschel/HIFI data of the CO (J=8-7) spectrum on Uranus compared to internal, cometary, and steady external infall source models (left), and the corresponding vertical profiles (right). The comparison of Herschel data to the CO source models allows for the inference of an external source of Uranus's CO.}
\end{figure*}

The detection of ``non-equilibrium'' C, N, O and S-bearing molecular species --- such as HCN, CO, and CS --- in the upper atmospheres of the giant planets has prompted questions pertaining to their origin and chemical evolution. Chemistry and circulation may result in the production and distribution of trace species in the upper atmosphere, as may the infall of cometary, interplanetary dust, or satellite material, or some combination thereof \citep{cav05, cav10, cav12b, cav13, cav14, cav19, lel10, iino20, tea22}. While previous observations at millimeter wavelengths have allowed for the investigation of these processes (Figure \ref{fig:ur_co}), higher sensitivity measurements are required to more fully characterize the governing sources and sinks of non-equilibrium species in the atmospheres of the giant planets. In addition, long-term observing campaigns allow for the measurement of temporal variability in these species, as well as potential sporadic events (e.g. storm outbreaks; cometary impacts) that may alter the upper atmospheric composition and can be measured by AtLAST. To perform these investigations requires wideband ($\gtrsim16$ GHz), high-resolution ($\lesssim1$~MHz) spectral imaging, as the spectral transitions from minor atmospheric constituents can manifest as significantly extended (several GHz wide), pressure broadened pseudo-continuum absorption wings around a narrow ($\sim$~MHz-wide) line core (see e.g., CO on Neptune; \cite{lus13a,lus13b}).

Finally, the detection of new, trace chemical species in the atmospheres of the giant planets will inform our understanding of their atmospheric photochemistry, and elucidate connections between the external sources from the surrounding system (e.g. dust, satellite infall) or beyond (e.g. comets, galactic cosmic rays). While ALMA has greatly facilitated the detection of heretofore unknown molecules in the atmosphere of Titan, the background continuum emission from the giant planets presents a challenge to the detection of all but the strongest emission lines (CO, HCN, CS, etc.) in their atmospheres. To further expand the known chemical inventory of molecular species, both predicted and serendipitous, the nominal dynamic range of AtLAST needs to be sufficiently high so that emission lines over 1000 times weaker than the continuum level can be reliably detected against large ($\sim20-30"$), bright (several hundred Kelvin) continuum sources. Chemical species of interest include HC$_3$N ($\sim136$, 264, 318, and 672 GHz), CH$_3$CN (bands at $\sim129$, 221, 312, and 698 GHz), CH$_3$CCH (bands at $\sim154$, 222, 325, 632 GHz), C$_3$H$_8$, and other nitriles and hydrocarbons. AtLAST therefore holds the promise of detecting new chemical species on giant planets through advancements in sensitivity and dynamic range, while a large focal plane array ($\gtrsim25\times25$ pixels in size) is required to map their spatial distributions without rotational smearing. Such observations would provide major insights into the origins of non-equilibrium species, through spatial associations with storms, aurorae or infall phenomena, and would lead to a new understanding of the extent of photochemical processing of gases in giant planetary atmospheres.

In addition, studies of molecular isotopic ratios provide profound insights into the formation and evolution of the giant planets (\cite{nom23} and references therein). The measurement of C-, N-, and O-bearing isotopes on the giant planets enables the determination of the main reservoirs for these species, providing crucial inputs for planet formation models. With a spectroscopic dynamic range $\gtrsim10^4$, measurement of more accurate isotopic ratios for $^{13}$C, $^{17}$O, $^{18}$O, and $^{15}$N in CO, HCN and other gases will be possible, resulting in improved understanding of how planetary systems form and evolve from the isotopically diverse ice, gas and dust reservoirs in the protosolar accretion disk.


\section*{Science Case ii: Terrestrial-Type Atmospheres}

Terrestrial-type planets with thick atmospheres are well suited to molecular spectroscopic studies at (sub)millimeter wavelengths. With atmospheric molecular column densities in the range $2.4\times10^{23}$--$1.5\times10^{27}$~cm$^{-2}$, and angular diameters $\sim1-60''$, Venus, Mars and Titan present compelling targets for spectroscopic studies using a sensitive (sub)millimeter facility. As shown in Figure \ref{fig:res} and Table \ref{tab:beams}, Mars and Venus can be readily resolved using a 50 m (sub)millimeter telescope, allowing spectral mapping of trace atmospheric gases to improve our understanding of their atmospheric physical and chemical processes.

\subsection*{Venus}

Mapping of spectral lines from Venus with strong transitions (such as SO$_2$, SO and CO) can provide insights into the chemistry and dynamics of our sister planet's atmosphere, much of which remains to be fully understood. The high spectral resolving power ($\sim10^6$--$10^7$) offered by heterodyne spectroscopy results in measurements mostly sensitive to altitudes $\gtrsim60$ km --- above the thick Venusian cloud layers --- allowing detailed studies of the mesosphere and above. Previous space missions to Venus including Pioneer Venus, Venera, Venus Express, and Akatsuki have allowed us to better understand the sulphur and water cycles and global circulation of the atmosphere \cite{mar18,tay18}. As described by \cite{enc19}, ground based studies provide a valuable complement to these datasets, thanks to their ability to monitor the behavior of gases as function of time, both in the short term ($\sim$hours) and the long term ($\sim$years or decades). With respect to space-based data, ground-based observations have the advantage of instantaneous snap-shot imaging over the whole disk of Venus, enabling global atmospheric studies as a function of latitude, longitude, and local time. 

The Venus Express and Akatsuki orbiters detected strong, long-term (multi-year) variations in the planet-scale zonal wind speeds and atmospheric SO$_2$ and CO abundances that remain to be fully understood. In the wake of these space missions, continued ground-based (sub)millimeter monitoring is therefore important to better understand these phenomena, which will help us understand the evolution of Venus's global climate. Spectral-spatial mapping of the entire ($\sim10-60''$ diameter) Venusian disk at a spectral resolution $\sim0.1$ km s$^{-1}$ and spatial resolution of $\sim$ a few arcseconds is required to fulfill this science objective.

The complex behaviour of sulphur-bearing molecules is a particularly important topic for further study due to their central role in Venus's climate, their close relationship with other volcanic gases (such as the ubiquitous clouds of sulfuric acid that blanket Venus), and their ability to trace photochemical and dynamical phenomena \cite{mar18}. The strong temporal and spatial variability of SO and SO$_2$ abundances at high altitudes remains largely mysterious, so further mapping studies are required to elucidate the sources and and sinks of these molecules in the upper atmosphere, and relate them to the lower altitude (volcanic) gases and aerosols. Such temporal studies are difficult using ALMA due to its variable array configuration that resolves out most of Venus's disk for a large part of the year, so rapid, multi-beam total-power mapping is required.

CO plays a key role in the carbon dioxide cycle, which maintains the chemical stability of the primary constituent of Venus’ atmosphere. Previous (sub)millimeter studies \cite{cla85,cla08,cla12,enc15} disagree on the extent of spatial variations in the CO distribution, so additional, temporally resolved mapping studies of this molecule are needed. In particular, future, temporally and spatially resolved, ground-based measurements of the atmospheric composition of Venus will be required to contextualize \textit{in situ} data from upcoming spacecraft missions such as DAVINCI+ \cite{gar22}, VERITAS \cite{cas21} and EnVision \cite{gha12}.

The unparalleled sensitivity of AtLAST would enable searches for new molecules (including sulfur, phosphorous and chlorine-bearing species, as well as organic molecules) that would help improve our understanding of the chemical processes occurring in the Venusian atmosphere, and their link to the volcanic gases that are thought to be so critical in governing Venus's hot global climate. Due to the bright (several hundred Kelvin) Venus continuum, and the comparative weakness of the mm-wave absorption lines from any yet-to-be-detected molecules ($\sim10$ mK), searches for new molecules will place strong demands on the spectroscopic dynamic range (DR) of the instrument (which would ideally reach DR $\sim10^5$ to allow in-depth studies of Venus's trace gas inventory). Close attention will therefore need to be paid to the bandpass calibration accuracy and spectroscopic baseline stability, in order to achieve very flat baselines, to allow us to realise the full potential of such a powerful telescope for planetary science.  A precise measurement or upper limit of the phosphine (PH$_3$) abundance would be just one of the ways in which AtLAST would help revolutionize Venus atmospheric science \cite{gre21,vil21,bai20,cor22}.

\begin{figure*}
\centering
\includegraphics[width=\textwidth]{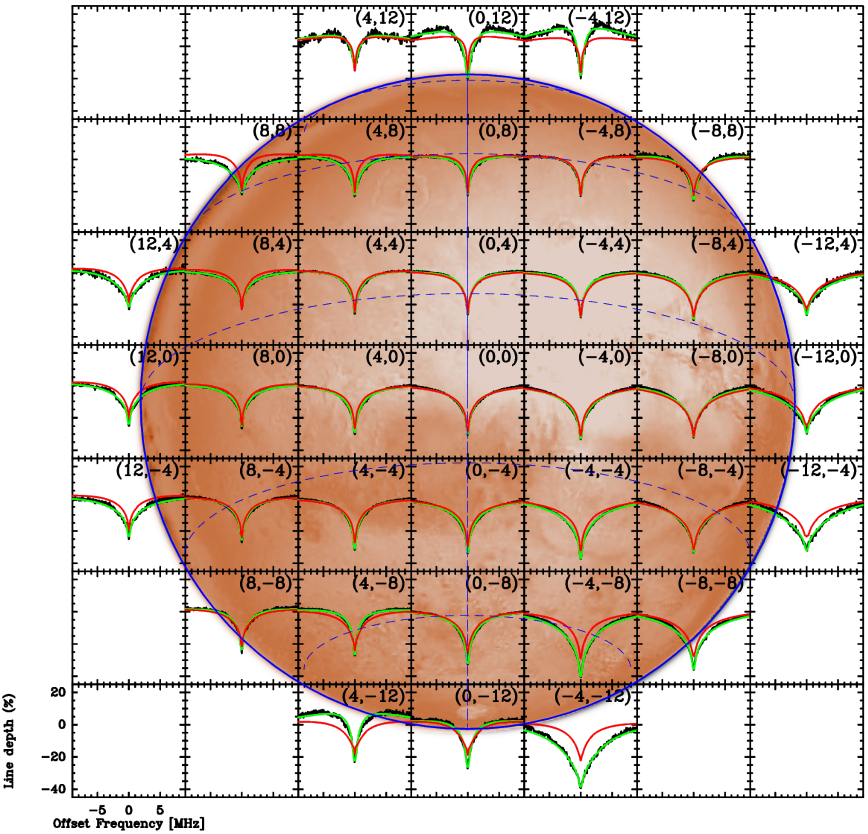}
\caption{\label{fig:mars} Spectral line map of CO $J=1-0$ absorption towards Mars, obtained using the IRAM Plateau de Bure Interferometer with a beam size $\sim7''$ (adapted from \cite{mor09}; overlaid on an optical map of the Martian surface from the NASA Solar System Simulator; https://space.jpl.nasa.gov/). Black lines: Observations. Red lines: Line profile expected from GCM predictions. Green lines: Fit of observations with retrieved thermal profiles.}
\end{figure*}

\subsection*{Mars}

The divergent histories of Earth and Mars contain valuable lessons in our quest to determine the conditions required for the origin and persistence of life on planetary surfaces. Understanding the present-day climate and atmospheric composition of Mars, and how it evolved over the history of the Solar System, is therefore an important objective in the fields of planetary science and astrobiology.  Mapping and detection of trace atmospheric gases such as CO and H$_2$O allows us to constrain photochemical networks and climate models, while measurements of molecular isotopic ratios provide unique insights into the planet's atmospheric loss history \cite{vil15,nom23}. Quantifying the Martian atmospheric circulation/global wind field is another essential objective towards revealing the past climate of Mars. 

Detailed studies of the composition and wind field of Mars's upper atmosphere were recently performed using data from the Mars Atmosphere and Volatile Evolution (MAVEN) spacecraft/orbiter \cite{ben19}. These measurements were obtained in the thermospheric altitude range ($\sim140$--240 km), using data acquired over several Martian seasons and spanning a large number of spacecraft orbits. Discrepancies between the observed wind speeds and those predicted by state-of-the-art climate models indicate that the global air circulation on Mars remains to be fully understood \cite{roe19}; further studies of these phenomena, including how the thermospheric wind speeds are influenced by those at lower altitudes, are therefore needed. Spatially complete measurements of Mars's 4''--26'' disk at high ($\sim0.1$ km s$^{-1}$) spectral resolution are required in order to derive the zonal wind speeds as a function of latitude, similar to those described for the giant planets (Science Case i).

The circulation in the middle atmosphere (40--80 km) is also affected by atmospheric transport and climatic processes in the lower atmosphere (\textless40 km). Although the lower atmospheric dynamics of Mars have been characterized by previous spacecraft missions (e.g. \cite{hin04,gra07}), and are reasonably well reproduced by general circulation models (GCMs), the middle atmosphere has been less well studied, so models are less constrained in this region \cite{mor09}. (Sub)millimeter heterodyne observations are uniquely well suited to probing the Martian middle atmosphere, yielding temperature and wind measurements from spectrally and spatially resolved molecular line profiles \cite{lel91,cav08}. As shown by \cite{mor09}, comparison of ground-based Mars CO mapping observations with GCM predictions (see Fig. \ref{fig:mars}) indicates significant discrepancies in the wind speeds and their spatial structure, and the GCMs often underestimate the temperatures in the middle (20--50 km) atmosphere.

No Mars space mission to-date has included high-resolution heterodyne instrumentation. Mars science would thus benefit from an improved-sensitivity ground-based (sub)millimeter telescope. With an angular resolution of a few arcseconds in the (sub)millimeter range (to cover e.g. the CO $J=3$--2, 6--5, 7--6 or 8--7 lines between 300--1000 GHz), combined with a large field of view, AtLAST would perform global monitoring of the middle atmosphere of Mars (over timescales from hours to years), enabling critical benchmarking and subsequent improvement of climate models.

\subsection*{Titan}

Since the end of the Cassini mission, ALMA has risen the forefront of ground-based Titan studies, thanks to its ability to perform spatial/spectral imaging of a large number of atmospheric gases, measure their (3D) distributions, and detect new species, including molecules of possible astrobiological relevance \cite{cor14b,cor15,cor19,pal17,nix20,the19a,the20}. This has demonstrated the power of long wavelength astronomy to provide new insights into the chemistry, dynamics, climate, and potential habitability of small icy moons in our Solar System.  Although a 50-m class facility such as AtLAST will not spatially resolve Titan (which has an angular diameter $\approx1''$, including its extended atmosphere; Table \ref{tab:beams}), with sufficient collecting area, bandwidth and atmospheric transmission, AtLAST would be able to compete with large (sub)millimeter interferometers such as ALMA and NOEMA in terms of flux sensitivity per beam. Consequently, broadband molecular line surveys with AtLAST would be able to reach unprecedented sensitivity towards objects such as Titan. Broad bandwidths (10's of GHz) have been shown to be crucial for the detection of increasingly complex organic molecules in extraterrestrial environments, due to their large number of individual rotational transitions spread across the cm/(sub)millimeter range \cite{biv15,mcg20,loo20,coo23}. Combining the fluxes from multiple emission lines in a broadband survey (e.g. between 200-300 GHz at sub-mK RMS and 1~km\,s$^{-1}$ spectral resolution), would therefore provide sufficient sensitivity to detect new molecules on Titan, including molecules of possible (pre-)biotic relevance, and help improve the accuracy of abundance ratio derivations.



Similarly, AtLAST could enable the detection and characterisation of new isotopologues of known molecules (such as $^{13}$C, $^{15}$N and deuterium-substituted forms of C, N and H-bearing organics), which have the power to elucidate the long-term physico-chemical evolution of Titan's mysterious atmosphere \cite{man09,man14,nom23}.

\section*{Science Case iii: Plumes from Icy Moons}

Advances in our knowledge regarding the interior structure of icy moons have demonstrated the likely presence of subsurface oceans throughout the Solar System \cite{nim16,she18}. Enceladus, Europa, Ganymede, Callisto and Triton are known or suspected ocean worlds. Cryovolcanic activity can launch plumes high into their atmospheres, and their molecular emission could be detected by a sensitive (sub)millimeter instrument. Plumes linking the subsurface with the atmosphere provide a potential means for probing the compositions of subsurface liquid reservoirs, and open up the possibility of remote habitability studies. Sensitive (sub)millimeter single-dish spectroscopy can probe the chemical compositions of the tenuous atmospheres of the icy Jovian satellites Europa, Ganymede and Callisto, the Saturnian satellite Enceladus, and the Neptunian satellite Triton to help determine the surface and/or interior compositions, from which insights into the chemistry and habitability of the subsurface oceans of these moons can be obtained.

Subsurface oceans are primary targets in the search for extraterrestrial life \cite{hen19}, but in-situ survey missions are extremely costly and difficult. One of the most outstanding results of the Cassini mission to the Saturnian system was the discovery of plumes emanating from the south polar region of Enceladus, believed to originate from a warm, subsurface ocean \cite{por06,spe18}. However, Cassini mass spectrometry of the plume was relatively low resolution leading to ambiguity in the detection of molecules with equivalent integer atomic mass (e.g. N$_2$/C$_2$H$_4$ and HCN/C$_2$H$_3$; \cite{wai09}). The possibility of chemical reactions and molecular dissociation within the Cassini mass spectrometer chamber is also difficult to quantify or rule out, as is the potential contamination of mass spectra due to Cassini fly-throughs of the Titan atmosphere. Unambiguous spectroscopic followup of the Enceladus plume composition is therefore essential. 
     
Following the end of the Cassini spacecraft mission, ground-based spectroscopic studies present a feasible alternative for studying the composition of the Enceladus plume(s), and complement observations from JWST \citep{vil23}. However, present observatories suffer from a lack of sufficient sensitivity and/or angular resolution to readily detect the faint, $\sim1''$ long plumes. Contamination from the bright, nearby ($<25''$ away) Saturn is also a major issue. Even large interferometers such as ALMA struggle due to the relatively large primary beam size provided by a 12~m sized antenna, which makes it difficult to exclude emission from Saturn and its rings from the field of view, leading to troublesome image artifacts. To avoid Saturn and exclude it from the 2nd and 3rd sidelobes of the primary beam, AtLAST would require a HPBW $<8''$ at 1.1~mm, and as such requires an aperture diameter \textgreater40 m.

As an example of what may be detectable, assuming a plume H$_2$O production rate of $\sim10^{28}$~s$^{-1}$, constant outflow velocity of 0.5 km\,s$^{-1}$, and rotational temperature of 25~K \cite{wai06,vil23}, with a jet opening angle of $30^{\circ}$ and HCN abundance of $10^{-4}$ ($10\times$ less than observed in typical comets \cite{boc17}), the peak antenna temperature for the HCN $J=4-3$ transition is calculated to be $\sim25$ mK (after convolution with a Gaussian beam of FWHM $3.5''$). Combining a small beam size with milli-Kelvin sensitivity, AtLAST would therefore be able to open up the possibility of direct ground-based studies of the Enceladus ocean chemistry.
        
Specific target lines for this science case include HCN (354 GHz), H$_2$CO (351 GHz), H$_2$O (183 GHz), CS (243 GHz), SO$_2$ (252 GHz), H$_2$S (369 GHz) and transitions of various salts and complex organic molecules including alcohols, carbon chains, aromatics and amino acids in the range 125--373 GHz. Detections (or useful upper limits) on the abundances of these species in the Enceladus and Europa plumes could thus be obtained. Searches for new molecules in the extended Enceladus torus \cite{har11} would also be possible. This would not only provide a crucial followup and validation of the Cassini mass spectrometry studies, but also allow changes in the plume composition as a function of time to be observed.

\section*{Science Case iv: Comets}\label{ref:case_iv}

Comets accreted in the Solar System at around the same time as the planets, and are believed to contain pristine (largely unprocessed) material from the protosolar accretion disk and prior interstellar cloud \cite{mum11}. Studies of cometary ices therefore provide unique information on the physical and chemical conditions prevalent during the earliest history of the Solar System. Due to their chemical compositions, rich in water and organic molecules, cometary impacts could also have been important for delivering the ingredients of life to otherwise barren planetary surfaces throughout the Solar System, and may therefore have played a role in initiating prebiotic chemistry \cite{chy92}. Examining the molecular and isotopic content of comets provide unique insights into the relationship between interstellar and planetary material.  Through broadband molecular surveys, we gain insights into the diversity of cometary compositions, improving our knowledge of the chemistry that occurred during planetary system formation. 

Based on prior experience with the CSO 10-m and IRAM 30-m telescopes \cite{boc00,biv14,biv16}, detections of new molecular species (previously undetected in comets), including complex organic molecules relevant to the origin of life, can be expected using a next-generation, single-dish (sub)millimeter facility. Observations of deuterated and $^{15}$N/$^{13}$C-substituted isotopologues of known molecules in a statistical sample of comets will provide fundamental new insights into the chemical origins of cometary (and thus, planetary) material. 

\begin{figure*}
\centering
\includegraphics[width=\textwidth]{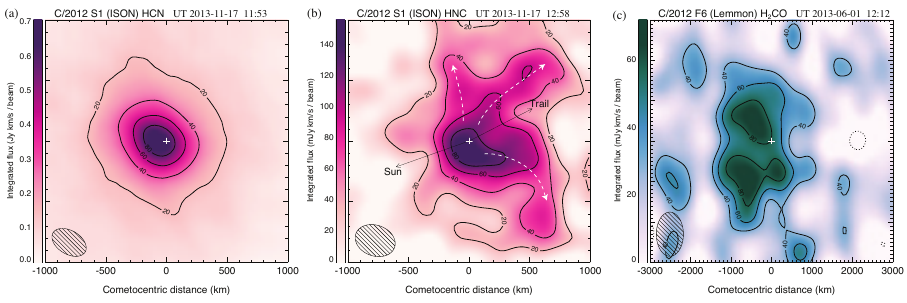}
\caption{\label{fig:comets} Contour maps of molecular line emission from in comets S1/ISON and F6/Lemmon using ALMA with a beam size of $\sim0.5''$. Contour intervals in each map are 20\% of the peak flux (the lowest, 20\% contour has been omitted from panel (c) for clarity). On panel (b), white dashed arrows indicate HNC streams/jets.  The peak position of the (simultaneously observed) 0.9~mm continuum is indicated with a white ‘+’. See \cite{cor14} for further details. Multi-beam mapping studies with AtLAST would dramatically improve our knowledge of cometary compositions and gas production processes on larger coma scales (up to several arcminutes).}
\end{figure*}

Spatial-spectral coma studies (see Figure \ref{fig:comets}) will provide fundamental information on the physics of cometary outgassing, as well as the complex thermal and photochemical processes occurring in the coma. Using currently-available telescopes, these crucial measurements can only be obtained in the very brightest comets (of which only a few appear every 10 years), so a more powerful (sub)millimeter facility (optimized for fast mapping) will significantly expand our knowledge and generate reliable statistics for the (chemically diverse) comet population.
        
\begin{figure*}
\centering
\includegraphics[width=\textwidth]{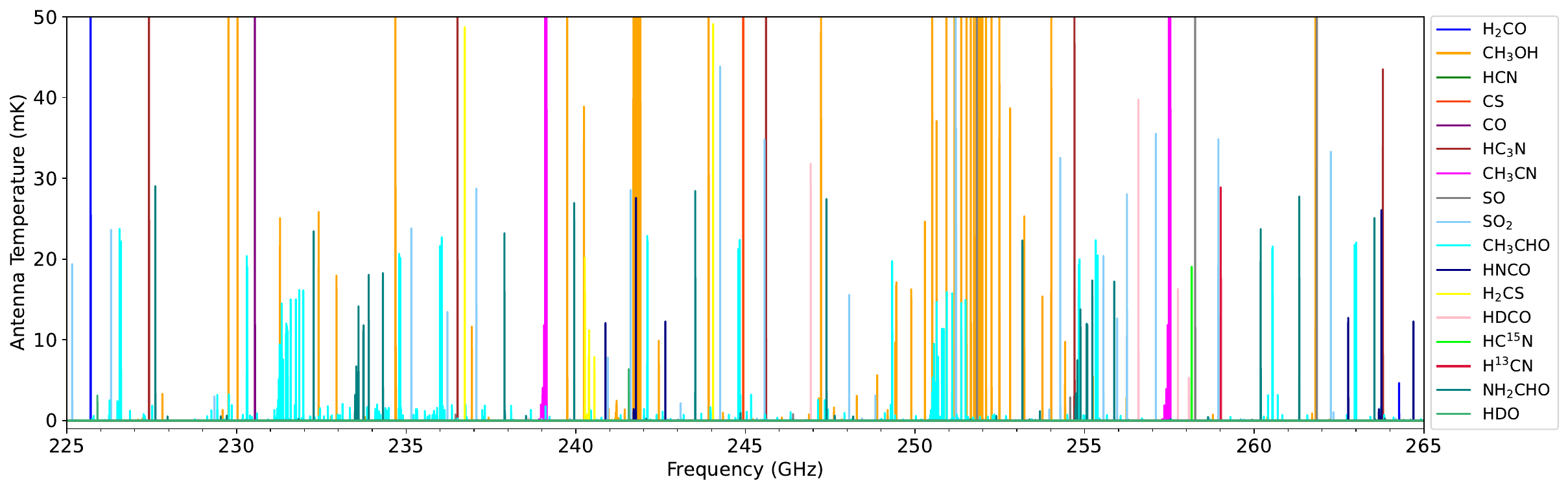}
\caption{\label{fig:cometlines} Broadband spectral model of a comet at 1 au from the Sun and Earth observed using a 50~m diameter telescope in the 1 mm band. We assume a typical cometary gas production rate of $Q({\rm H_2O}) = 10^{29}$ s$^{-1}$, spherically symmetric outflow velocity of 0.8 km\, s$^{-1}$ and rotational temperature of 60~K. Molecular abundances are the average of previously observed cometary values \cite{del16,boc17}.}
\end{figure*}

Specific target lines include HCN (354 GHz), DCN (362 GHz), HC$^{15}$N (344 GHz), H$_2$CO (351 GHz), HDCO (335 GHz), D$_2$CO (342 GHz), H$_2$O (183 GHz), HDO (465 GHz, 894 GHz), D$_2$O (607 GHz, 898 GHz) and transitions of various complex organic molecules including alcohols, carbon chains, aromatics and amino acids in the range 125-373 GHz (see Figure \ref{fig:cometlines} for a selection of the available lines from known cometary species in the millimeter waveband).

\begin{figure*}
\centering
\includegraphics[width=0.5\textwidth]{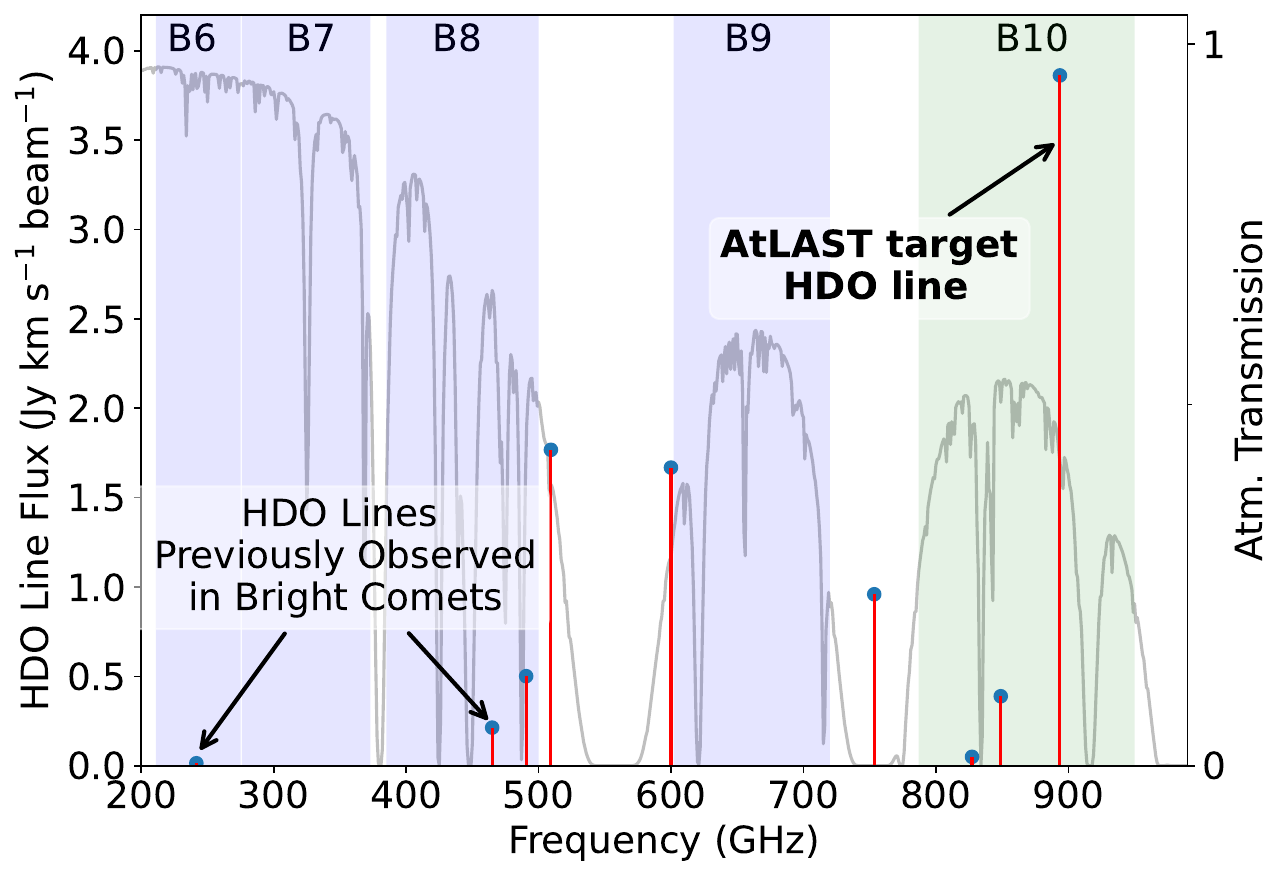}
\raisebox{5mm}{\includegraphics[width=0.37\textwidth]{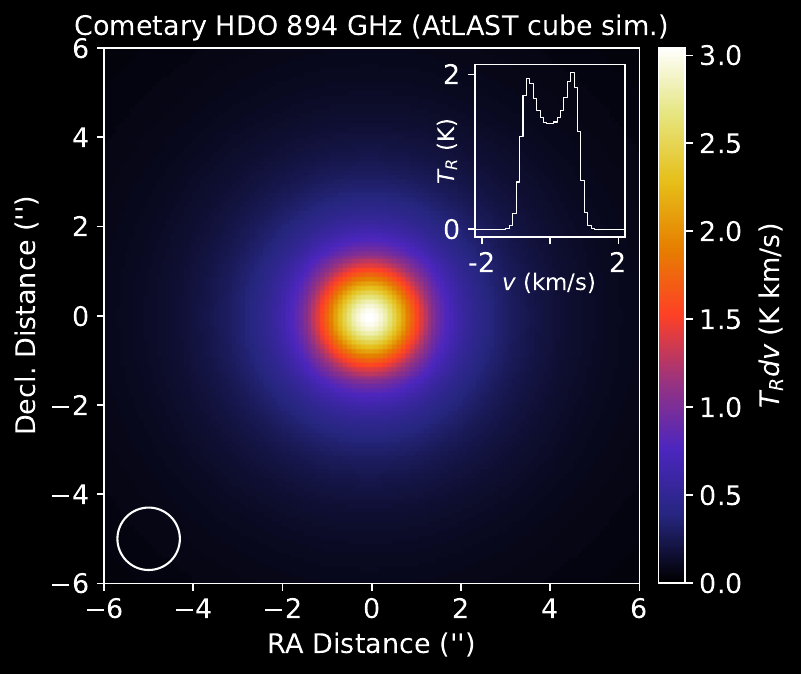}}
\caption{\label{fig:comethdo} Left: Simulated HDO line strengths in a typical (moderately bright; $Q({\rm H_2O}) = 10^{29}$ s$^{-1}$) comet at 1 au from the Earth and Sun, with average cometary HDO/H$_2$O ratio \cite{pag17}, using a diffraction-limited 50~m telescope beam size. The strongest HDO line in AtLAST’s frequency range is highlighted, which is much stronger than the cometary HDO lines observed previously in the (sub)millimeter band. Right: Simulated HDO 894 GHz map and spectral extract from the central pixel. The $1.4''$ AtLAST beam FWHM is shown lower left.}
\end{figure*}

Isotopic studies provide a powerful means for understanding the evolution, origins and inter-relationships between different Solar System bodies. In particular, materials throughout the Solar System exhibit differing degrees of deuterium (heavy hydrogen) enrichment, the measurement of which can reveal their chemical and thermal histories. Comets are thought to contain a significant amount of pristine, deuterium-enriched interstellar material, predating the birth of the Solar System \cite{alt17}. However, D-enrichment in water ice may also be possible due to fractionation reactions in the protosolar accretion disk, before the gas dissipates \cite{nom23}. The fraction of pristine (interstellar) vs. reprocessed (disk) ices incorporated into planetary materials is an important open question that can be addressed via improved measurements of cometary D/H ratios. In particular, comparison of cometary D/H ratios with protoplanetary disk models can help unravel the details of where, and from which materials cometary ices were accreted \cite{wil09,alb14,cle14}. As shown in Figure \ref{fig:comethdo}, the strongest cometary HDO line accessible using ground-based sub-mm telescopes is the $1_{1,1}-0_{0,0}$ line at 894 GHz. The next strongest lines at 509 and 600 GHz are on the flanks of the high-opacity 557 GHz H$_2$O line, making them undesirable targets, so the 465 GHz line presents the next best alternative after the 894 GHz line. A simulated HDO 894 GHz flux map (and spectral line profile) for a typical comet (assuming HPBW = $1.4''$) is shown in the right panel of Figure \ref{fig:comethdo}. Access to this line would provide an unprecedented opportunity to survey HDO in a large number of comets (several per year), spanning different levels (and modes) of outgassing activity, which would provide new insights into the origin of water on Earth, and elsewhere in the Solar System \cite{har11,cle14,alt15,nom23}.

Due to their spatially extended nature (up to several hundred arcseconds), and $\sim1/r$ brightness profile, optimal sensitivity to weak cometary lines is achieved using a large, single-aperture antenna rather than an interferometer, since the latter is blind to coma structures larger than the angular scale probed by the shortest baselines. To expand the statistics for cometary abundance measurements in a meaningful way (observing 2 moderately bright comets per year over a 10-year timespan), with sensitivity to all the species in Figure \ref{fig:comets}, will require milli-Kelvin sensitivity across $\sim1$~km\,s$^{-1}$ spectral line width. Since comets are dynamic, time-variable objects (on timescales of days or even hours), simultaneous observation of the species of interest is desirable for reliable molecular abundance comparisons. An instantaneous bandwidth of several 10's of GHz is therefore required, at a spectral resolution $\sim0.1$~km\,s$^{-1}$ in order to resolve the complex kinematical structure of the coma (e.g. \cite{rot21}). Simultaneous, multi-chroic observations in several frequency bands are required to probe the complete range of species of interest.  The ability to rapidly respond to transient cometary phenomena such as outbursts or surprise apparitions will necessitate flexible telescope scheduling, for example, with a turnaround time of a few days between identification of the target phenomenon and acquisition of the observations. It will also be important to accommodate time-constrained observing requests (ideally, to an accuracy of up to a few minutes), to perform synergistic observations with ground or space-based observatories at other wavelengths, or to capture specific, temporally isolated events such as rapid perihelion/perigee passages or spacecraft flybys.

In terms of frequency coverage, sensitive access to the (sub)millimeter range is required, ideally up to and including the HDO $1_{1,1}-0_{0,0}$ line at 894 GHz line.

\section*{Conclusion}

We have highlighted some the most important planetary science use cases for a new, large-aperture, single-dish submillimeter telescope. In particular, the improvements in spectral line sensitivity and dynamic range offered by such a cutting-edge facility will permit the detection of new molecules and isotopologues that will directly enhance our understanding of chemical complexity and abundances in the upper atmospheres of the terrestrial and giant planets, as well as in comets. In turn, these observations will improve our understanding of exoplanetary atmospheric dynamics and compositions, and allow for the improved assessment of planets and moons within our solar system to harbor past or present habitable environments. Instantaneous (snapshot), high resolution, broadband spectral-spatial molecular mapping will provide new insights into transient phenomena such as planetary winds and storms, as well as jets, outbursts and ambient molecular distributions in cometary comae and icy moon exospheres. The relevant instrument specifications required to achieve this science are summarized below.

\section*{Summary of Instrument Requirements}

\subsection*{Sensitivity}
RMS sensitivities of $\sim1$ mK per 1~km\,s$^{-1}$ spectral resolution element in a few hours on-source will be required for spectroscopic characterization of Solar System bodies (which are often time-variable). This will allow spectral mapping of known trace species, isotopic measurements, and new molecular detections in planetary atmospheres and comets.

\subsection*{Spectroscopic Dynamic Range}
Spectroscopic dynamic range (ratio of continuum to line intensity) $\gtrsim10^5$ is required for detection of weak atmospheric spectral lines from trace gases, against the bright continuum of Venus, Mars, Jupiter, Saturn, Uranus, Neptune and Titan. 

\subsection*{Spectral Coverage and Resolution}
(Sub)millimeter spectroscopic observations of planetary and cometary atmospheres require a combination of broad bandwidth (several GHz or more to properly characterize the profiles of strongly pressure-broadened rotational lines) and high spectral resolution ($\sim100$ kHz to fully resolve thermally-broadened line profiles $\sim1$ km\,s$^{-1}$ wide). Broad (10's of GHz) of bandwidth in the (sub)millimeter range --- preferably, simultaneously across multiple receiver bands --- is also necessary to observe multiple molecules and rotational transitions simultaneously, and would enable spectral line stacking analyses for more complex molecules whose lines are often distributed over a large frequency range. This will lead to orders-of-magnitude improvements in sensitivity for the more complex molecules, which can have hundreds of lines within the spectral range of interest. Instantaneous bandwidth of 32--64 GHz would be desirable to avoid having to change receiver tunings during time-critical observations, facilitating relative abundance and isotopic ratio analyses. 

\subsection*{Spatial Resolution}
A small beam size of $\sim5''$ at 300 GHz is required for the majority of planets in our Solar System to be spatially resolved (see Table \ref{tab:beams}), paving the way toward more detailed atmospheric studies than have previously been possible with single-dish facilities. Such a small beam is also required to minimize beam dilution for the small ($\sim0.1''$--$2''$ diameter) bodies such as icy moons and their plumes/exospheres. A small primary beam ($\lesssim8''$) is also strictly required for Enceladus observations, to avoid contamination due to its close proximity ($<25''$) from the bright emission sources of Saturn and its rings. 

\subsection*{Mapping Requirements}
Spatial resolution of a few arcseconds over a field of view up to a few arcminutes, with $\sim20$--50 beams across a single axis (for example, in a square or hexagonal array, $\sim25$ pixels in across) is required for snapshot mapping of terrestrial and gas-giant planetary atmospheres. A 50-m single-dish telescope with large focal plane array could achieve this without resolving out the larger planets, which is a problem for interferometers.

\subsection*{Time-domain Considerations}
Observations of Solar System objects are often time-critical, in the case of observing transient atmospheric phenomena, cometary outbursts etc., and to account for moons being in and out of eclipse/occultation, or to avoid observing when a satellite is too close to its parent planet. Target-of-opportunity observations, coordination with other telescopes/space missions, and accommodation for regular monitoring of science targets with a predefined cadence (over a timescale of days, weeks, months or years), will all be crucial for maximum science return on time-variable Solar System objects. Scheduling the most critical observations to within an accuracy of a few minutes will therefore be desirable. Responding to unexpected transient phenomena such as cometary outbursts or surprise apparitions will necessitate flexible telescope scheduling, for example, with a turnaround time of a few days.

\section*{Data and Software Availability}

The calculations used to derive integration times for this paper were done using the AtLAST sensitivity calculator, a deliverable of Horizon 2020 research project `Towards AtLAST', and available from \href{https://github.com/ukatc/AtLAST_sensitivity_calculator}{GitHub}. Spectral line intensities for Enceladus and comets were calculated using the \href{https://github.com/lime-rt/lime}{LIME} and \href{https://github.com/mcordiner/sublime-d1dc}{SUBLIME} radiative transfer codes, respectively \cite{bri10,cor22b}, both of which are available on GitHub.

\section*{Competing Interests}
The authors declare no competing interests.

\section*{Acknowledgements}
This work received support from NASA’s Solar System Obervations (SSO) Program, National Science Foundation (NSF) grant \#AST-2009253, NASA's Planetary Science Division Internal Scientist Funding Program through the Fundamental Laboratory Research work package (FLaRe), the UK Science and Technology Facilities Council grant ST/Y000676/1, and the European Union’s Horizon 2020 research and innovation programme under grant agreement No. 951815 (AtLAST). LDM acknowledges support by the French government, through the UCA\textsuperscript{J.E.D.I.} Investments in the Future project managed by the National Research Agency (ANR) with the reference number ANR-15-IDEX-01. ML acknowledges support from the European Union’s Horizon Europe research and innovation programme under the Marie Sk\l odowska-Curie grant agreement No. 101107795. SW acknowledges support by the Research Council of Norway through the EMISSA project (project number 286853) and the Centres of Excellence scheme, project number 262622 (``Rosseland Centre for Solar Physics''). 


{\small\bibliographystyle{apj_mod}
\bibliography{refs}}

\end{document}